# Leveraging reconfigurable micro-resonator soliton crystals for Intensity-Modulated Direct Detection Data Transmission

X. X. Chia[1,*], K. Y. K. Ong[1,*], A. Aadhi[1], G. F. R. Chen[1], J. W. Choi[1], B.-U. Sohn[1], A. Chowdury[1], and D. T. H. Tan[1,2,^]

1. Photonics Devices and Systems Group, Engineering Product Development, at the Singapore University of Technology and Design, 8 Somapah Road, Singapore 487372
2. Institute of Microelectronics, Agency for Science, Technology and Research (A*STAR), 2 Fusionopolis Way, Singapore 138634

*Authors contributed equally
^ dawn_tan@sutd.edu.sg

Abstract: The perennial demand for highly efficient short-haul communications is evidenced by a sustained explosion of growth in data center infrastructure that is predicted to continue for the foreseeable future. In these relatively compact networks, cost-sensitivity is of particular importance, which limits options to direct detection schemes that are more cost efficient than their coherent counterparts.  Since their initial demonstration, multi-soliton states in optical microresonators have been observed to manifest in self-organised ensembles where soliton pulses are equally spaced around the resonators. In the spectral domain, these states – dubbed soliton crystals (SCs) – result in significant enhancements to individual comb lines depending on the crystal state, making them well suited towards intensity-modulated direct detection (IMDD) schemes. In this work, we experimentally demonstrate adiabatic, deterministic access to lower-order soliton crystal states using an auxiliary-assisted cavity pumping method, attaining up to 19.6 dB enhancement of the comb lines in the 7-SC configuration compared to the single-soliton state. Seven comb lines of each 46 Gbaud/s pulse amplitude modulation 4 (PAM4) is transmitted over 4km of fiber in comb lines across the C-band with bit-error-rates (BER) as low as 5E-5. Our demonstration shows the promising way of using soliton crystal states as future integrated sources for highly stable Terabaud/s datacenter communications.

## 1. Introduction

The advent of high-Q microresonators in the last decade enabled the realisation of soliton microcombs[1,2] that remain one of the most intensely researched topics in nonlinear integrated optics. A single microresonator generating multiple wavelengths of light through parametric processes has the potential to eliminate the need for a multitude of bulky coherent laser sources. Therefore, amongst the many applications in which they have been employed (including optical clocks[3,4], spectroscopy sources[5], and rangefinders[6]), their potential as sources for high-capacity, long-haul communications are of significance, with data transfer on the scale of hundreds of terabits on bright soliton combs[7,8] and petabit communications on dark-pulse-kerr combs[9] being demonstrated in recent years. While these communications schemes are highly promising for long-haul coherent networks particularly in telecommunications, they tend to be unsuitable for a compact infrastructure for intra– or inter-data center interconnects due to system complexity, phase noise, synchronization and higher cost sensitivity which results in a preference towards intensity modulated direct detection (IMDD) encoding schemes such as non-return to zero (NRZ) and Pulse Amplitude Modulation (PAM) formats.

In data center communications, the bright single-soliton (SS) combs that have been the source of choice for long-haul communications suffer due to their relatively lower power and efficiency, compounded by the relatively higher detection floors of receivers used in direct detection. Fujii et al[10] demonstrated a bright SS comb with 10 GHz line spacing for direct detection schemes across 40km of fiber, but data rates were limited at 10 Gb/s NRZ, with error-free transmission occurring only on the 2 lines adjacent to the pump and BERs of $10^{-3}$ to $10^{-5}$ reported on the rest across the C-band, attributing the relatively poorer BERs to low comb line powers. In addition, the aforementioned work utilizes a non-monolithic, non-CMOS compatible $MgF_2$ whispering gallery mode microresonator. Platicon combs fare much better due to greatly enhanced comb line powers at the expense of bandwidth, and have been demonstrated to transmit PAM4 data across 2km of optical fiber on a normal-dispersion AlGaAsOI resonator at BERs between $1\times10^{-2}$ to $5\times10^{-4}$ with 32 Gbaud/s and $2\times10^{-2}$ to $1\times10^{-3}$ to with 50Gbaud/s[11].

Even though soliton combs are a promising solution for ultrahigh-capacity transmission links through massively parallel wavelength channels, the low-power drawbacks of bright SS combs and limited bandwidth in platicons restrict access to more comb lines. A new class of soliton, called bright soliton crystals (SCs) are promising candidates to overcome these challenges. Particularly, these self-organised ensembles of soliton pulses provide significant enhancement to individual comb lines depending on the N-SC state (where N indicates the number of pulses circulating in the cavity) while retaining extremely robust and stable operation along with high spectral efficiency and spectral bandwidth of bright soliton combs. In $\chi^{(3)}$ cavities such as silicon nitride and silica microresonators, the generation of SCs often leverages resonances with avoided-mode-crossings (AMXs)[12,13], exploiting the localized dispersion of the high intermodal coupling to anchor the SC state. As communications sources, their feasibility has also been demonstrated in long-haul coherent communications[14] and shorter fiber reaches[15], with the latter demonstrating 22 Gbaud/s PAM4 on comb lines across the L-band, attaining BERs around $10^{-3}$ over 10km of fiber using a distributed feedback diode (driven) driven, self-injection locked silicon nitride microresonator.

Thus far, the use of SCs in $\chi^{(3)}$ microresonators as transmission sources involve having to account for the stochastic nature of accessing lower order soliton crystal states[12,16], which limits its integration in practical communications applications. This is largely attributed to the thermal effects that hinder deterministic access, which have the added effect of pinning the attainable N-SC state on the geometry of the rings. While configurable soliton crystals have been demonstrated by leveraging the backscattered pump light to induce beating within the cavity against the auxiliary laser[16], it is noteworthy that other works in which the auxiliary-assisted method was employed for soliton generation did not report similar observations of soliton crystals and access to lower order states remained stochastic[17]. Previously, 2-PSC states using the auxiliary-assisted generation method were reported where control over the relative positions of the circulating soliton pulses in the cavity could be manipulated by changing the positions of the pump laser relative to the AMX.[18]

In this work, we demonstrate the generation of perfect soliton crystals (PSCs) with high extinction ratios between the main SC lines, yielding up to 19.6 dB amplification per line in the 7-SC configuration when compared to single-soliton combs. By leveraging the auxiliary-assisted method to provide adiabatic access to the soliton states, we find that straddling the AMX resonance by placing the pump and auxiliary lasers symmetrically around the resonance allows for the reliable generation of PSCs from 3-SC state to the 7-SC state and high repeatability (>90%) for the 2-SC state, extending results from earlier works and adding a highly beneficial layer of reconfigurability to fit into existing communications channels. In addition, similar to observations in[18], we find that the relative locations of the circulating soliton pulses can be manipulated by coupling the wavelength of the auxiliary laser to the adjacent resonance which resolves suspected interference issues while retaining the high comb-line amplification in the main SC lines (in this work henceforth labelled as the imperfect soliton crystal (ISC) state for clarity), albeit with slight power penalties between 0.1-2dB compared to the PSC. For the first time, we demonstrate IMDD data transmission (NRZ and PAM4) using a reconfigurable PSC and ISC source from a CMOS device. We successfully encode up to 46 Gbaud/s data on 7 comb lines across the C-band in the ISC state for an aggregate 322 Gbaud/s, achieving bit-error-rates on the order of $10^{-5}$ both in the direct configuration and across 4km of optical fiber, well within the forward error correction (FEC) limit.

## 2. Soliton Crystal Generation

A micro-ring resonator fabricated on silicon-nitride ($Si_3N_4$) is employed for the generation of the frequency combs. The resonator is designed with a ring-radius of 100 μm and a waveguide cross section of 1.5 μm × 0.8 μm, yielding a free-spectral-range (FSR) of 227 GHz (~1.8 nm) (Fig. 1a) and anomalous dispersion across the telecommunications bands easily observed from the downward trend of the FSR (in GHz) with increasing frequency. From the transmission spectrum, a large reduction in the extinction ratio and double-peak pattern is observed at the resonance located at 1545.5 nm in both the $TE_{00}$ and $TM_{00}$ modes, as a result of strong mode coupling between the fundamental modes of both polarizations indicative of an avoided mode crossing (AMX). The dispersive change by the avoided mode crossings creates an extended background wave which interacts with the pump laser and auxiliary laser. This subsequently arranges the multi-solitons that are propagating in the cavity whereby the solitons interact with each other via their respective extended oscillatory tails [13] which has been successfully employed for the generation of soliton crystals using a single-pump laser in the past. The symmetrical positions of the pump and auxiliary laser thus generates the formation of perfect soliton crystals deterministically.

We generate bright-soliton frequency combs from a high-Q microring resonator (Fig. 1b), using the auxiliary-assisted method[19], wherein light is coupled into the resonators from opposing ends of a bus-waveguide using fiber circulators (Fig. 1c). In this configuration, the auxiliary laser aids in thermally stabilising the cavity, allowing for an adiabatic transition from chaotic modulational instability (MI) combs to the soliton states and enabling consistent access to the microresonator soliton regime. This method has also been observed to generate perfect 2-SC[18] combs by pumping away from the AMX resonance, with variations in the pump laser location allowing for control over the relative azimuthal location of the 2 circulating pulses.

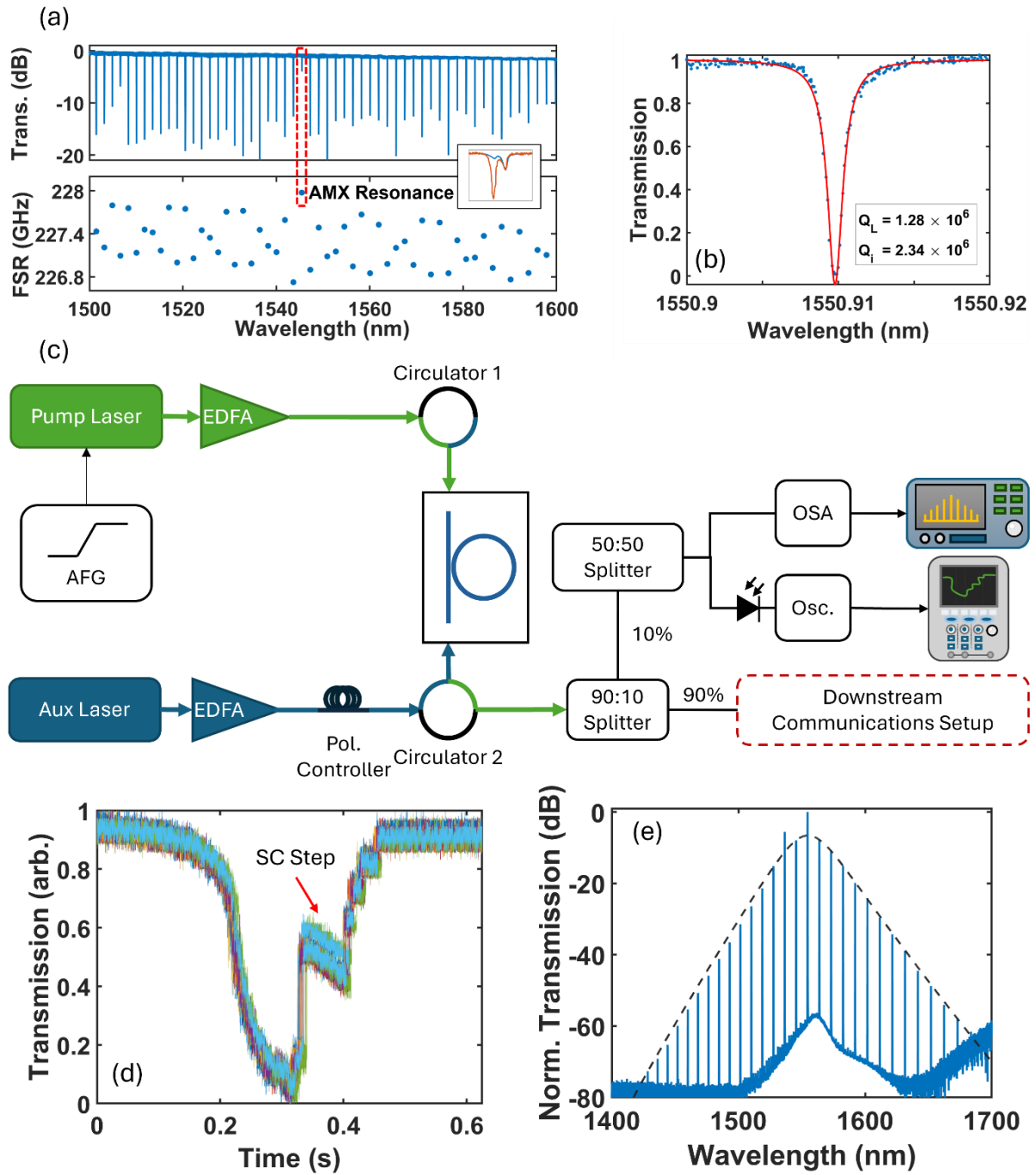

Figure 1 (a) $TE_{00}$ transmission spectrum and accompanying FSR of the micro-ring resonator. A large reduction in the extinction ratio of the resonance and local variation in the FSR is observed at 1545.45 nm, corresponding to a resonance with high intermodal coupling between the $TE_{00}$ and $TM_{00}$ modes. The inset shows the transmission spectrum of the resonance in the $TE_{00}$ (—, blue line) and $TM_{00}$ (—, orange line) modes, showing good overlap in both polarizations, indicative of an avoided mode crossing (AMX). (b) Measured spectra (•, blue circles) and Lorentzian fit (—, red line) of a resonance located at 1550.9 nm (c) Experimental setup for soliton generation and data experiments. (AFG: Arbitrary Function Generator, EDFA: Erbium-Doped Fiber Amplifier, OSA: Optical Spectrum Analyzer, Osc.: Oscilloscope) (d) Overlay of 20 consecutive oscilloscope traces in the 5-SC configuration as the pump is swept across the resonance, showing consistent repeatability in generating the soliton-crystal step for the soliton generation scheme (e) Corresponding 5-SC transmission spectrum spectra (—, blue line) and $sech^2$ fit (--, black dotted line) from the 5-SC step shown in 1(d)

For the N-SC generation, in our experiment, we employ on-chip pump and auxiliary powers of 80 mW and 105 mW respectively. By straddling the AMX resonance with the pump and auxiliary sources placed symmetrical N-FSRs apart from the AMX resonance, an N-SC microcomb can be reliably generated (full details of the generation method along with traces are described in Supplementary Materials Section I), yielding soliton crystals from 2-SC to 7-SC states depending on the wavelength of the laser source. Multiple steps in Figure 1(d), corresponding to various soliton states, are clearly observed in the oscilloscope trace even without filtering the pump and auxiliary lasers, with the very first and last step of each trace always corresponding to the N-SC state and single-soliton state respectively. Figure 1(e) also shows a close $sech^2$ fit of the 5-PSC spectra, one of the key indicators of microresonator DKS formation. Since the limited tunability of the AMX resonance limits our ability to filter out the pump and auxiliary sources, we rely on dips in the transmitted power trace to determine when the various soliton states are accessed, as opposed to the usual comb power traces. Figure 1(d) shows a plot of 20 overlaid consecutive power traces when the pump is swept over the resonance in the 5-PSC configuration, showing steps corresponding to the various soliton states wherein the N-SC and single-soliton states occur every time the pump laser is scanned across the resonance.

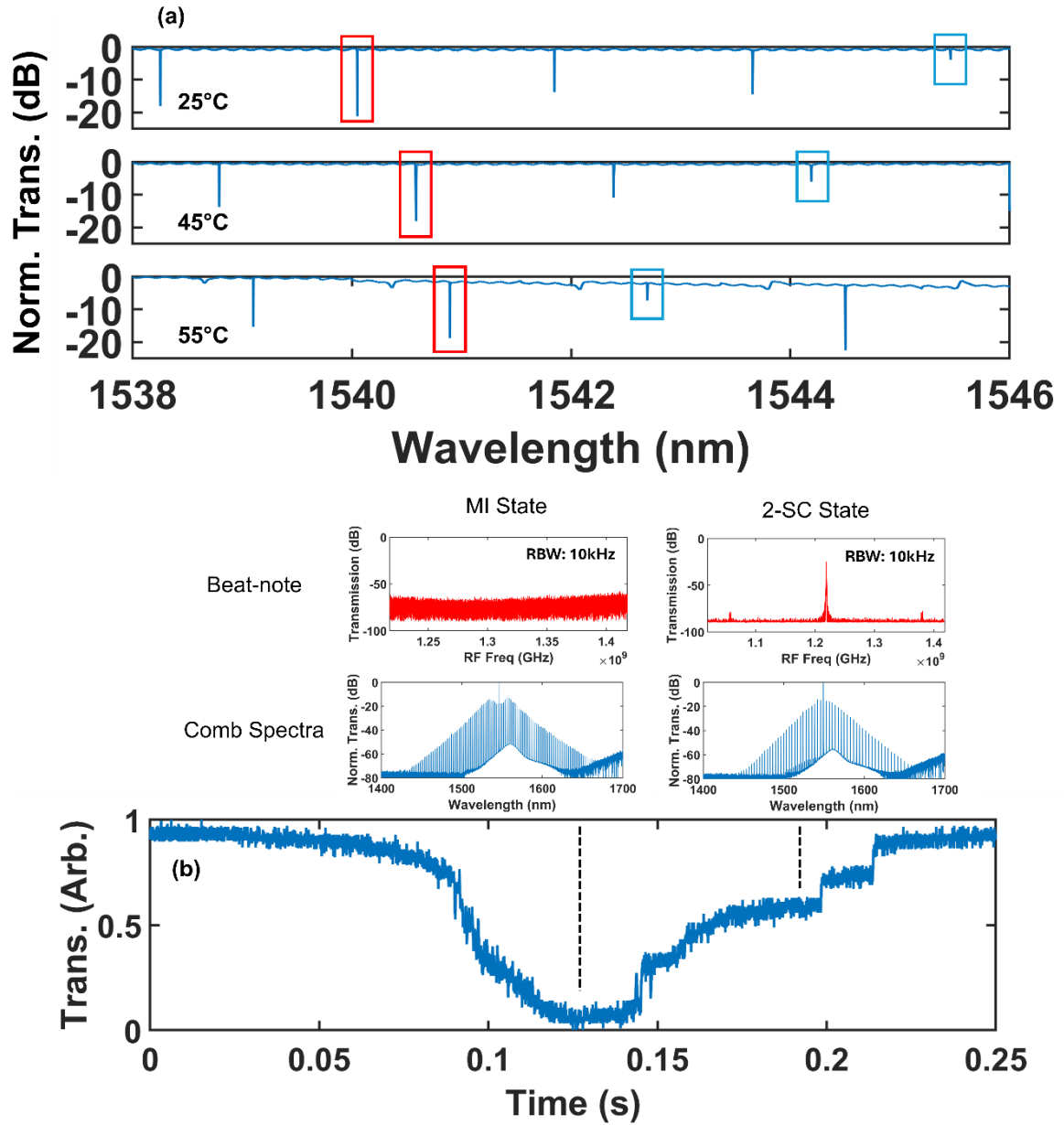

Figure 2 Thermal AMX tuning and beat-note characterisation experiments (a) TE transmission spectrum of the ring resonator at various temperatures on the TEC. A red shift of the native $TE_{00}$ resonances (red boxes, ☐) and blue-shift of the AMX resonance (cyan boxes, ☐ ) is observed with increasing temperature, attributed to the higher temperature sensitivity and higher FSR of the $TM_{00}$ modes that result in them coinciding at shorter wavelengths with increasing temperature (b) 2-SC oscilloscope trace as the pump laser is swept across the resonance from the blue to red side. Corresponding comb states (blue line comb spectra, —) and RF beat-note (red line RF spectra, —) measured from the real-time oscilloscope are shown above the trace. Black dotted lines (--) indicate the location of the trace at which the comb states are generated. Smaller side-band peaks in he beat-note plot of the 2-SC state is attributed to the co-polarised auxiliary laser and has also been observed in[19]

To confirm the coherence of the generated state as proof of soliton formation, we perform beat-note experiments by beating a generated comb line in the PSC state with a low-noise distributed-feedback diode (DFB) laser at a wavelength of 1550.02 nm. Since the room-temperature location of the AMX resonance at 1545.45nm and ring FSR of 1.8 nm limits PSC line formation near the DFB laser frequency, the resonator needs to be heated with a thermo-electric cooler (TEC) temperature controller to ensure the generated lines and the laser wavelength coincide. While heating the resonator results in a red-shift of the native TE and TM resonances, the higher group index associated with the TM mode undergoing a larger change in wavelength compared to the TE modes, leading to a blue-shift of the AMX resonance to 1542.8 nm instead (as shown in Fig. 2a). This allows for the generation of a 2-SC comb with a comb line at 1550 nm. For beat-note visualisation, a comb line from a SC is combined with the DFB laser using a 50:50 coupler and detected by a high-speed photodiode. The resulting beat signal is then observed in a real-time oscilloscope. Fast-Fourier Transform (FFT) of the measured beat-note at a resolution bandwidth (RBW) of 10 kHz shows significant noise reduction when the comb is tuned from the MI state to the 2-SC state, confirming good coherence and soliton formation. We note that the co-polarized pump and auxiliary lasers coupled simultaneously into the cavity result in the formation of comparatively broader sidebands around the beat-note peak (Fig. 2b), an observation previously highlighted in[19] that is of importance for high-speed communications experiments in the following section.

In addition to coinciding with soliton steps in the power trace, the spectral envelope of both N-SC and single-soliton combs shows a clear sech$^2$ characteristic pattern of microresonator soliton states[12,19,20]. A full set of comb-trace overlays and 2-SC to 7-SC combs successfully generated from the ring resonator is shown in section I of the Supplementary Materials. When attempted without the use of an auxiliary laser (i.e. with only the pump laser[12]), only the 7-SC combs could be repeatably accessed, 6-SC combs occurred stochastically, and 5-SC combs and below could not be accessed at all even with lower pump powers. The soliton crystal states are accompanied by significant enhancements in power/comb-line, ranging from 17.8 dB to 19.6 dB enhancement in the 7-SC state compared to the single-soliton comb. For 4-SC combs and above, the comb lines are observed to exceed 1 mW across the C-band after off-chip coupling, making them highly suitable for IMDD communications applications.

# 3. IMDD Transmission

## 3A. Reconfigurable Soliton Crystals

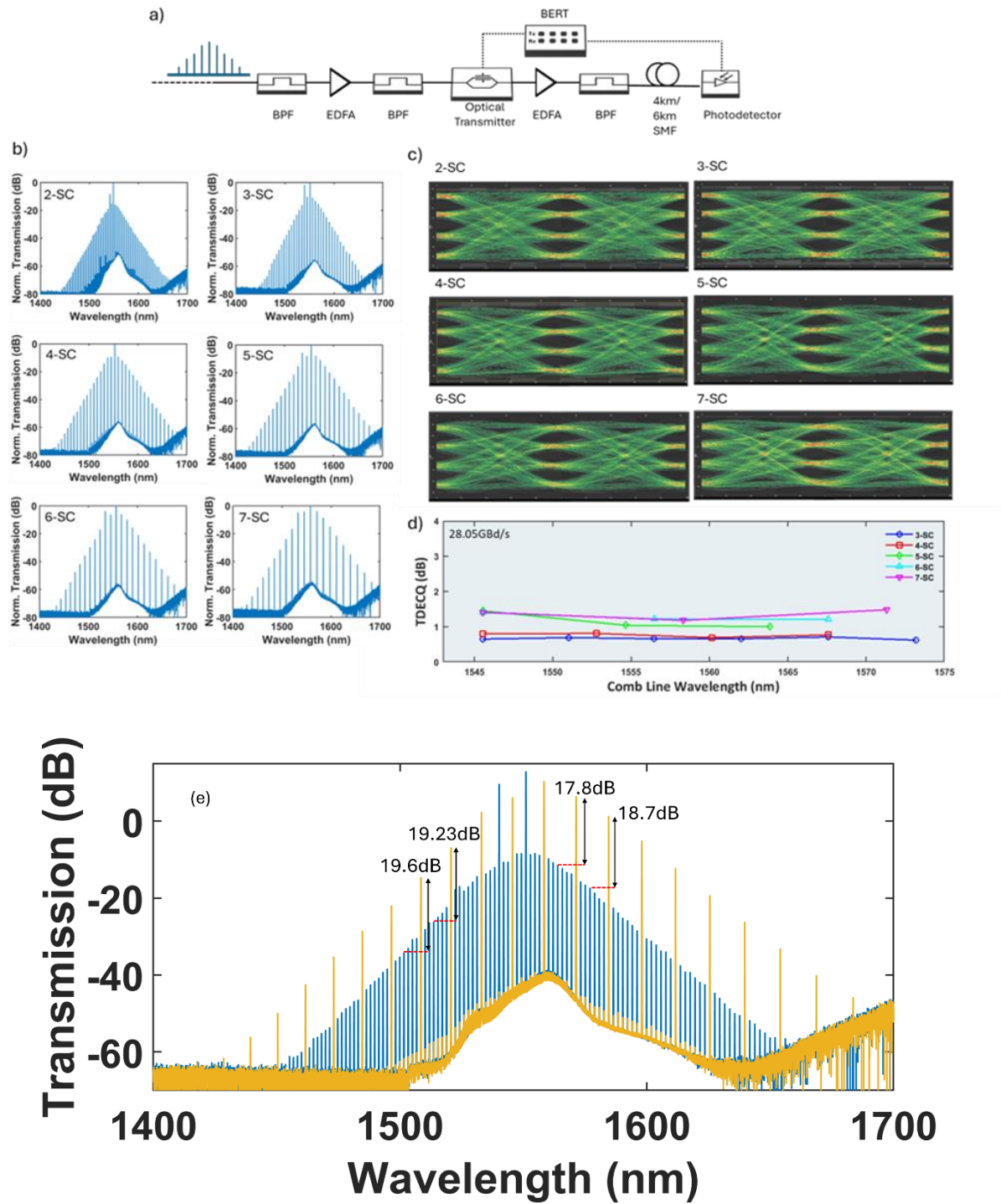

Fig. 3 High-speed measurements using reconfigurable soliton crystals. 3 a) Schematic of the high-speed measurements. b) Spectra and c) Eye diagrams of the respective N-soliton crystal states for comb lines adjacent to the pump line. d) TDECQ measurements from the transmission of PAM4 28.05GBaud/s for various soliton crystal (N-SC) states. e) Optical power difference between a 7-SC and a single soliton. PSC

comb lines are compared to comb lines in the single-soliton regime based on wavelength separation from the pump (BPF – Band Pass Filter, EDFA – Erbium-doped Fiber Amplifier, BERT – Bit-error Rate Tester)

We begin by investigating the efficiency of the PSCs as sources for direct detection schemes. The comb lines are individually filtered and amplified before being encoded with 28.05 Gbaud/s NRZ-OOK and 46 Gbaud/s PAM4 data. Their performance is measured (1) directly from the transmitter (back-to-back) and (2) through 4km of single mode fiber. The performance is quantified using the transmitter dispersion eye closure quaternary (TDECQ) and bit-error-rate (BER). Both systems used in tandem are highly useful as the BER provides information about the quality of the transmission while the TDECQ is a metric used to assess PAM4 data with lower values (dB) indicating better transmitter performance.

Figure 3a shows a schematic of the transmission setup while Fig. 3b shows the spectrum for multiple N-SCs with the comb line separation fixed by the symmetrical FSR spacing between the auxiliary and pump laser with the AMX resonance. The ability to reconfigure the perfect soliton crystal number is advantageous for wavelength division multiplexing (WDM) parallel data transmission. It allows for flexibility in accessing the soliton crystal comb lines and comb-line power across wavelength channels. Furthermore, the N-SCs can be generated deterministically (See Supplementary Section S1). The application of soliton crystals in communication channels would thus require low soliton crystal numbers, whereby a larger FSR would minimize the available number of channels in a communication band. The eye diagrams for the various N-SCs are shown in Fig. 3c, attaining open eye diagrams for comb lines directly adjacent to the pump line. It can also be observed that the eye openings are slightly skewed[21] with an increase in the soliton number. This also correlates to an increase in the TDECQ with the increase in the soliton state number as shown in Fig. 3d. Lastly, comparing to the microcombs of a single soliton, the carrier to noise ratio of soliton crystals is greater. This is evidently shown in Fig. 3e where the optical power of the comb lines in the C-band of a 7-SC is greater than a single soliton by 17.8dB to 19.6dB.

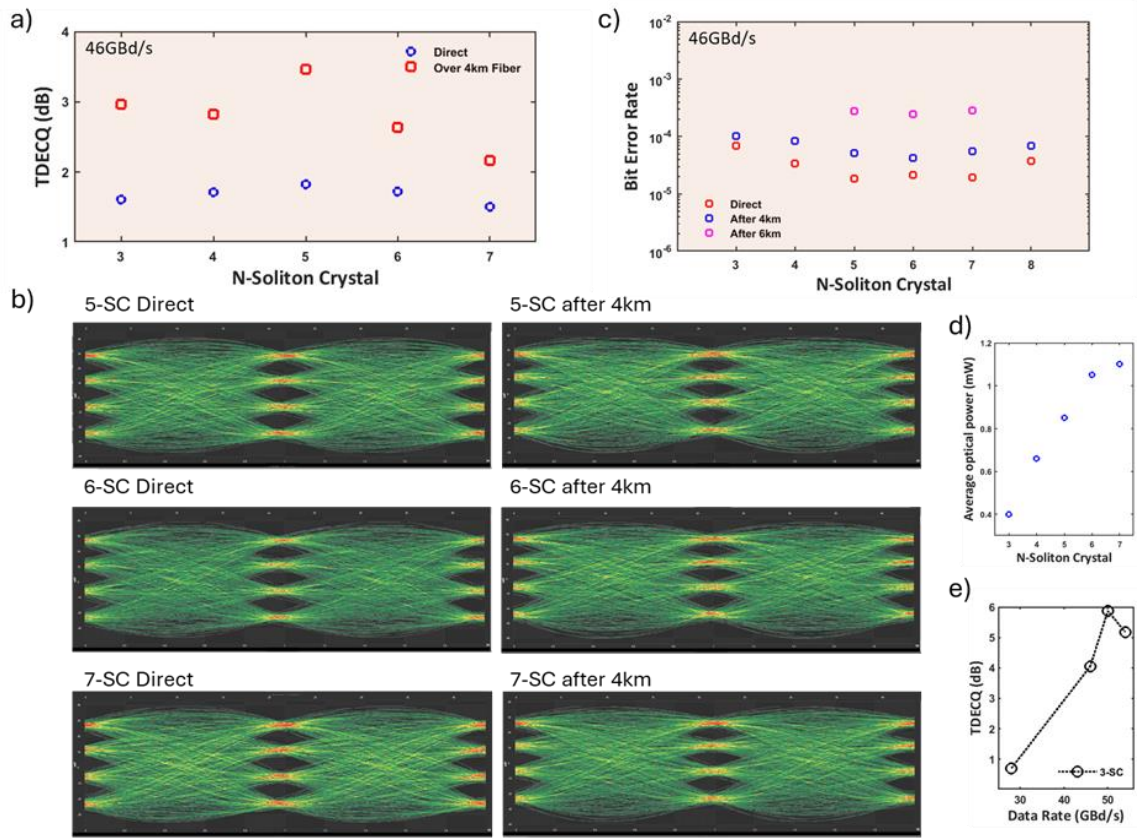

Fig. 4. TDECQ characterization. a) TDECQ measurements as a function of the soliton crystal number (N-SC) with PAM4 46GBaud/s and b) eye diagrams respectively for a direct configuration and after 4km of SMF. c) BER measurements as a function of N-SC for a direct configuration, over 4km and 6km of SMF. d) The average optical power as a function of N-SC before data modulation. e) TDECQ measurements for various data rates.

The high-speed data transmission links using N-soliton crystal comb lines are characterized with TDECQ and BER measurements, as shown in Fig. 4. The comb lines at the AMX is modulated with 46GBaud/s PAM4 data and all measurements are performed at the same optical power from the transmitter of 1.5 dBm. Figure 4a shows a penalty of ~1dB in the TDECQ over 4km of single-mode fiber (SMF) as compared to a direct (back-to-back) configuration, with an eye opening for both configurations shown in Fig. 4b. The BER measurement for a direct configuration and after propagation through 4km of SMF is shown in Fig. 4c. Both TDECQ and BER show a similar trend in values across various N-SCs where TDECQ and BER values are <3.5dB and <$10^{-4}$ respectively. Across the N-SCs, as shown in Fig. 4d, the average optical power of the AMX comb line increases with the soliton number but Figs. 4a and c show that there is no significant difference in the TDECQ and BERs. Furthermore, to investigate the bit rate independent performance of the system, we operated it with the data rates ranging from 28.05 GBaud/s to 54 GBaud/s, as shown in Fig. 4e. The TDECQ and BER increase with data rates as expected. The consistent TDECQ values and BERs across the N-SCs indicate that reconfiguring across the N-SCs does not affect the performance of the communication channels.

## 3B. High-Speed Data Transmission of perfect soliton crystals (PSC)

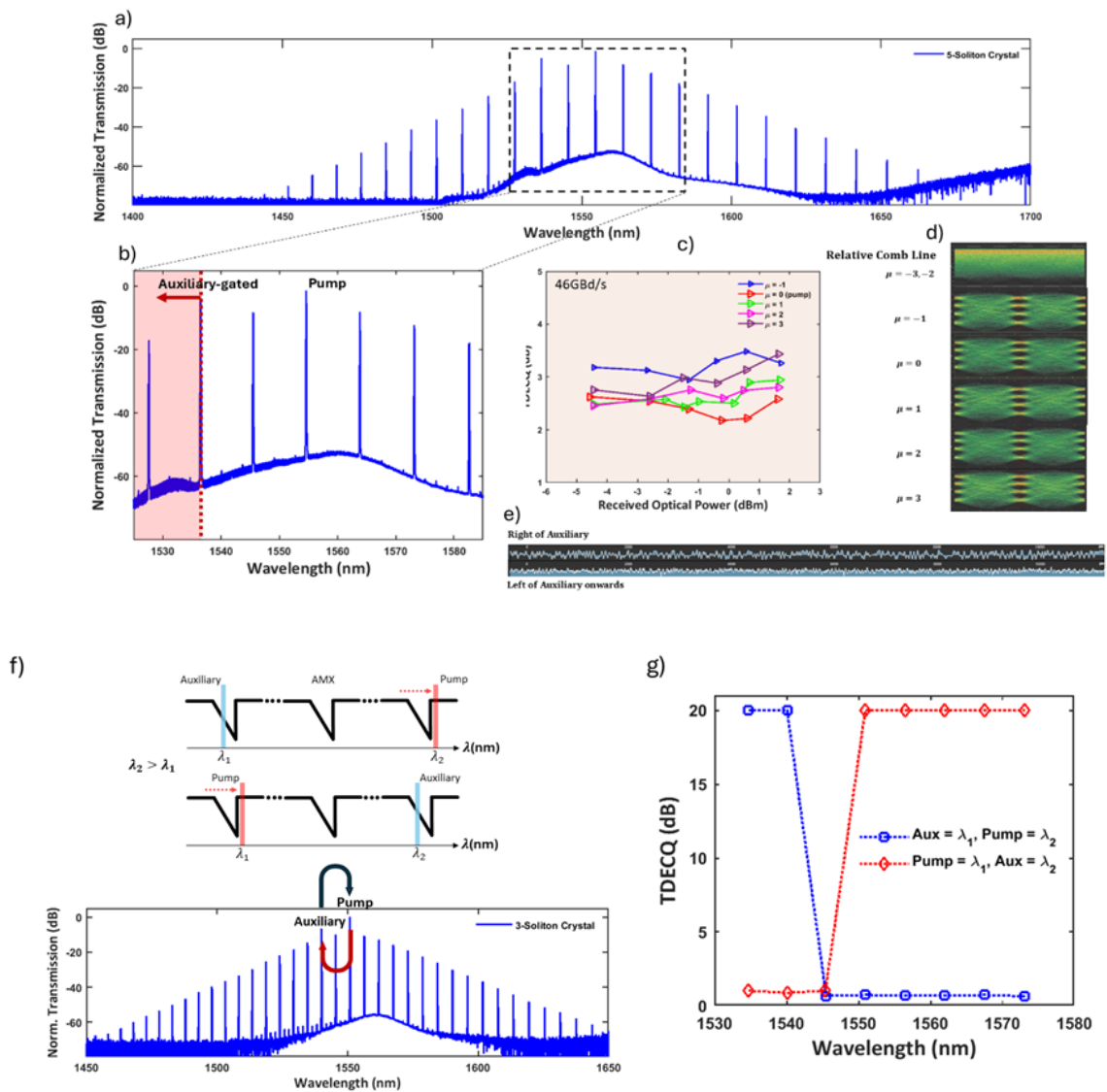

Fig. 5 High-speed measurements where the pump and auxiliary lasers are within the C-band. a) Measured spectrum of the 5-Soliton Crystal with b) microcombs within the C-band and the Auxiliary-gated region (highlighted in red). c) TDECQ as a function of the received optical power over 4km for the C-band comb lines on the right of the Auxiliary-gated region. d) Eye diagrams for the respective comb lines indicating a contrast in data transmission. e) The sample waveforms for the comb lines that are modulated with PAM4 46GBaud/s data and the Auxiliary-gated region. f) Optical spectra of a 5-SC with a swap of the pump and auxiliary laser position with respect to the AMX resonance. g) TDECQ as a function of the comb lines for a 5-SC with the swapping configuration.

The number of comb lines that can be used for data transmission is limited by the resonance wavelength of the auxiliary laser employed to achieve the perfect soliton crystal state. It induces noise for shorter (longer) wavelengths where the auxiliary resonance wavelength is shorter (longer) than the pump resonance wavelength. We termed this as auxiliary-gated data transmission highlighted in red in Fig. 5b, where data transmission of the comb lines with a fixed number of FFE taps throughout the C-band coincides with a region that has a significant amount of noise. The TDECQ values from the AMX comb line was measured as a function of the received optical power as shown in Fig 5c. The transmission of PAM4 46GBaud/s data over 4km of SMF shows TDECQ values <3.5 dB. The eye diagrams in Fig. 5d show the contrast between the comb lines below and above the auxiliary-gated region. We have repeatedly observed this effect in other N-SCs, where an open eye diagram is clearly visible from the AMX mode (between the pump and auxiliary modes). The contrast in the waveforms between the two regions is also shown in the sampling waveform, in Fig. 5e, of PAM4 modulated data across the sampling time. As evident from the figure, a four-level intensity modulated waveform can be observed as compared to a noisy waveform with an auxiliary-gated comb line. Our investigation suggests that due to interference from the auxiliary laser, data transmission is observed only on the red side of the auxiliary laser wavelength. On the contrary, when the auxiliary and pump laser positions are switched, as shown in the diagram in Fig. 5f, the opposite effect occurs in the generated SC where data transmission is observed only on the blue side of the auxiliary laser wavelength. The measured TDECQ as shown in Fig. 5g, with corresponding open eye diagram for TDECQ <2dB and closed eye diagram for a maximum TDECQ value of 20dB. If the position of the auxiliary laser is at a shorter wavelength than the pump laser, the first two frequency channels obtain a maximum TDECQ value of 20dB with an optimum TDECQ value from the AMX mode. Upon switching positions of the pump and auxiliary lasers, the opposite is observed. Despite limiting the number of 'usable' microcombs, it facilitates switching the wavelength required for data transmission.

The phenomenon could be explained by the forced oscillations as predicted and demonstrated in recent work, where soliton pulse vibrates around its fixed position[16,22] Influenced by the counterpropagating auxiliary laser, the force acting on the soliton translates to an XPM-induced sideband generation around the auxiliary comb line. Several other works have also reported the translation and generation of microcombs around the auxiliary laser with extended spectral bandwidth in different spectral bands[23,24]. The formation of a secondary comb shown in Fig. 6a at the auxiliary output, that share the same repetition rate, is mainly depends on the coupling between the soliton crystal microcombs and backscattering of the soliton pulse[25], and the power ratio between the counterpropagating pumps[26,27]. Figure 6a shows the optical spectrum of the 5-SC at the pump output (shown in blue) and the counter-propagating pulse at the auxiliary output (overlaid in green). The slight skewing effect observed in the eye diagrams in Fig. 3c arises from the extent of chirp and phase modulation due to the dependence of refractive index change[28-31] on the soliton number and the relative power increase in the microcomb. Our experimental observations in the high-speed measurements conclude that the phase noise from the XPM induced mode detuning is greater in the auxiliary gated region, resulting in high TDECQ values and closed eye diagrams. This effect is more prominent near the auxiliary mode and is interestingly biased towards shorter or longer wavelengths, depending on the location of the auxiliary and pump laser wavelengths. Furthermore, the power ratio between the microcombs of the soliton crystal and the auxiliary generated microcombs determines the extent of nonlinear interaction (SPM and XPM) and thus affects the amount of phase noise contributing to this wavelength dependency.

## 3C. High-speed data transmission with imperfect soliton crystals (ISC)

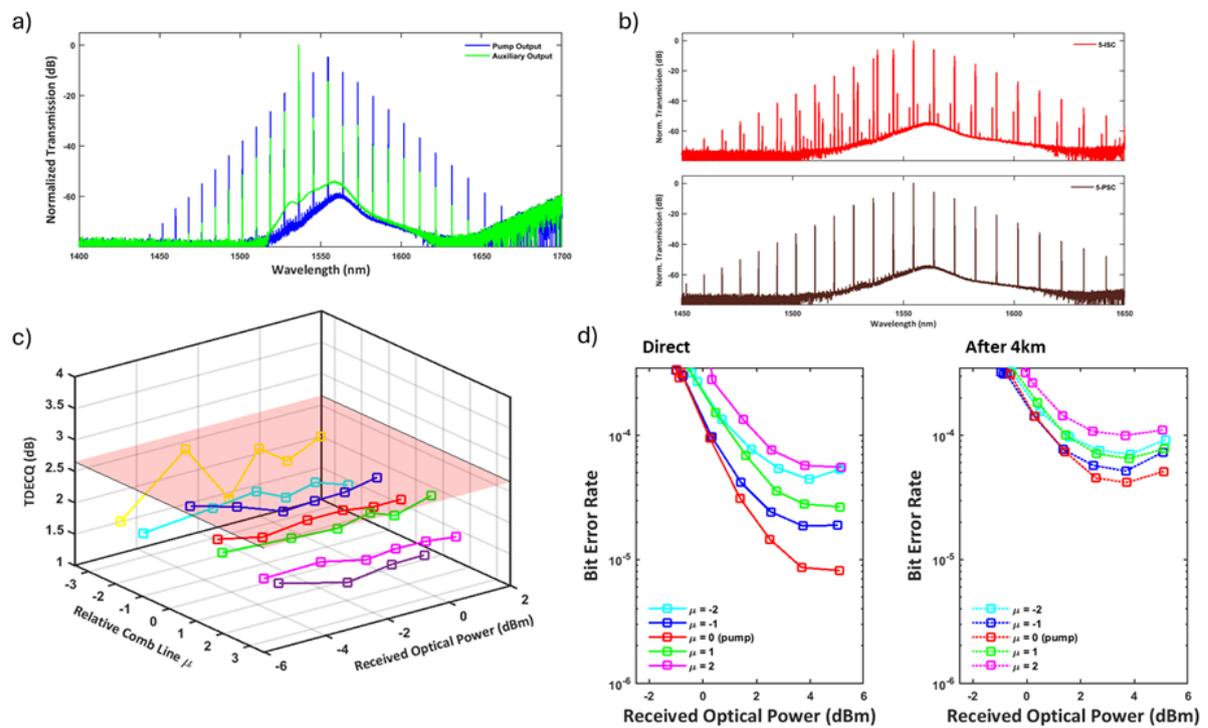

Fig. 6 High speed data transmission with imperfect soliton crystals. Optical spectra of a) the 5-SC at the pump output (blue) overlaid with the auxiliary output (green) and b) the 5-imperfect soliton crystal (top) and 5-PSC (bottom). c) TDECQ as a function of the received optical power across the comb lines in the C-band with the 5-ISC. d) BER measurements as a function of the received optical power for the direct configuration and after 4km of SMF.

In the dichromatic pumping scheme, tuning the position of a pump with respect to the auxiliary wavelength is akin to shifting the azimuthal position of the soliton in the resonator cavity. This forms a new and separate type of soliton crystals, namely imperfect soliton crystal (ISC), also known as a soliton crystal with defects[13]. In our setup, the symmetrical FSR spacing between the pump and auxiliary lasers with respect to the AMX resonance is disrupted by varying the auxiliary laser position by one or more FSR. By increasing the FSR spacing between the pump and AMX resonances[18], it has also been shown to decrease the relative azimuthal positions of the solitons within the cavity. As shown in the top panel of Fig. 6b, a 5-ISC is generated by shifting the auxiliary laser by one FSR towards the AMX. Furthermore, we investigate the suitability of ISC for high-speed data transmission. Figure 6c shows TDECQ values to be <2.6dB across the comb lines after PAM4 modulation at the data rate of PAM4 46GBaud/s, demonstrating data transmission inclusive of the auxiliary gated region across the comb lines in the C-band, which was previously hindered by the auxiliary laser. The shift in the auxiliary laser by one FSR generates additional sidebands within the soliton crystal states due to crystal defects, as seen in a 5-SC microcombs (see Fig. 6c). This results in the XPM generated sidebands shifting away from the microcombs of the PSC used for data transmission. The measured BERs across the comb lines lie within the range of $3 \times 10^{-5}$ to $2 \times 10^{-4}$, as shown in Fig. 6d over 4km of SMF, meeting the standard BER thresholds of $1.5 \times 10^{-2}$ and $4.5 \times 10^{-3}$ for soft-decision FEC (20% overhead) and hard-decision FEC (7% overhead) respectively[32]. A comparison of the BERs of the PSC microcomb and adjacent sideband is shown in the supplementary. The lowest BERs are observed at the pump comb line and the BERs increase as the comb lines move further away from the pump, showing a similar trend over 4km.

## 4. Discussion and conclusion

Modern communication systems are moving towards regional bases of data centers and inter-data centers connections that require cheaper yet robust solutions for data transmission, and soliton crystals are very promising as optical sources for multiple wavelengths across a wide bandwidth. Their main advantage over single solitons is their superior transmission comb power and stability — the average optical power of individual comb lines is not only distributed across the pulse energy of the soliton, but the lower detuning range also results in higher coupling of the pump laser to the resonator cavities. As such, soliton crystal pulses present with higher pulse powers on top of the increased repetition rate, which leads to a quadratic relationship between the comb power per comb line increasing and the soliton number ($\propto N^2$), leading to higher optical signal to noise ratios (OSNR) and greater efficiencies when used as a source for optical transceivers. Similar comb power to soliton crystal number relationships have been observed in other works featuring soliton crystals in $\chi^{(3)}$ cavities[12,13,16], which evidences an empirical relationship that can be further explored rather than an idiosyncratic observation.

Prior work reporting bright soliton crystal combs for high-speed communications has been shown to generate comb lines with limited power and data formats. On the other hand, dark pulse Kerr combs circumvent this power limitation due to higher conversion efficiencies and have been more suitable for PAM4-based encoding schemes, with recent reports demonstrating higher comb line powers of > -10dBm within the C-band[33,34]. When compared to this work, however, BERs have been observed to be comparatively higher[11]. We note also that in our demonstration, the on-chip power from N-SC states across the C-band is > 2mW (>1mW off-chip), which is considerably higher than usually achieved using bright solitons and is comparable to DPK combs. The high power in each of the comb lines is of great significance, since the key barrier to the adoption of frequency combs in transceivers is the limited lane power. Notably, CWDM4[35] and PSM4[36] transceivers used in datacenter communications serve reaches up to 2km and 500m respectively, requiring a minimum single lane power of -6.5dBm and -9.4dBm respectively. 100GBASE-LR4 transceivers standardized by the IEEE serve a 10km reach, and stipulate a minimum lane power of -4.3dBm[37]. Importantly, the comb lines demonstrated in Fig. 3e and 6b exceed the lane power requirements for all these standards, firmly placing it within practical applicability in commercial datacom hardware.

In addition to comb power enhancements, another key advantage of reconfigurability is the relaxation of fabrication and design constraints for the combs to align with the existing communications standard channels. In the past, this would be intimately tied to the geometric design of the resonator given that comb line spacings are limited by the FSR of ring designs. While significant advances have been made in microresonator DKS generation over the years, reliable generation of solitons has thus far been inherently geometry dependent. Adiabatic access to the single-soliton state using a single pump laser requires rings with THz FSRs[38,39] which exceed ITU grid spacing and limits comb output power due to the smaller interaction length and bending losses present in small ring radii. Soliton crystals also suffer from similar limitations, where accessing the SC state using single-pump schemes is limited to a specific pump power window and in the context of $\chi^{(3)}$ cavities, AMX to pump laser spacing[12,13,38]. Therefore, the minimum achievable N-SC is also intimately tied to ring geometry[32] as adiabatic access requires a minimum intracavity power to overcome thermal effects in spatiotemporal chaos to SC state transitions. Although variations in resonance tuning and pump powers have shown some measure of reconfigurability in the SCs such as in[32], the configuration window is ultimately limited and all the demonstrated states have a line separation of ~1THz, far exceeding the grid spacing.

The reliable access to reconfigurable, low-order soliton states and slight tuneability of the AMX resonance in our work allows for the relaxation of fabrication and design constraints of the ring resonators since the spacing between comb lines and their location can be directly manipulated. With the dual-pump configuration, the AMX location can be engineered outside of the C-band and subsequently adjusted through deft placement of the auxiliary and pump wavelengths to maximize the number of comb lines with effectual data transmission[39-41]. An added consideration is that higher soliton numbers offer greater optical microcomb power, but lower soliton numbers are preferred to maximize the number of comb lines that can be used within the communication band, and configurability would be greatly beneficial to tuning the resonator to suit various application requirements. The added flexibility in source reconfigurability and control of the comb lines through thermal tuning ensures that the generated comb lines align with WDM channel locations. Additionally, the ease of implementing such a frequency comb system with controlled channel spacing and optical power via the pump and auxiliary laser positions reduces the complexity of operation as well as increases the potential for integration into optical transceivers for IMDD data transmission.

The combination of dichromatic pumping and utilizing the avoided mode crossing allows direct access to PSC as well as manipulation of the relative pulse positions to create controlled defects we have termed imperfect soliton crystals. The auxiliary laser stabilizes the thermal effects upon red detuning of the pump laser, but introduces XPM-induced effects that generates phase noise in several comb lines in the PSC configuration, leading to high BER values. However, this can be mitigated with the use of imperfect soliton crystals, which largely retains the amplification of key comb lines while mitigating the induced phase noise.

In conclusion, we have demonstrated deterministic generation of various perfect soliton crystal states with a symmetrical placement of the pump and auxiliary lasers with the AMX resonance as a pivot. IMDD data transmission of the soliton crystal comb lines modulated with 46GBaud/s PAM4 data over 4km of SMF within the C-band, and the measured BERs lie within the range of $3 \times 10^{-5}$ to $2 \times 10^{-4}$, exceeding the BER thresholds of SD- and HD-FEC schemes. The reconfigurability in the repetition rate and comb line spacing, aided by the FSR spacing of the pump and auxiliary resonance with respect to the AMX, allows for the tunability of the comb lines to fit various WDM channels. A solution to overcome the auxiliary laser-induced phase noise in the PSC combs is presented by shifting the wavelength of the auxiliary laser. Lastly, the enhanced averaged optical power in the soliton crystal combs increases the viability of implementing the optical transceivers for communication systems that require a cheaper alternative, and robust generation of WDM channel sources.

## Ethics Declarations

The authors declare no competing interests.

## Acknowledgements

Funding from the A*STAR MTC Grant (M21K2c0119), National Research Foundation Investigatorship (NRF-NRFI08-2022-0003), Ministry of Education ACRF Tier 2 Grant (T2EP50121-0019), Quantum Engineering Programme 2.0 grant (NRF2022-QEP2-01-P08) and A*STAR Institute of Microelectronics (C220415015) is gratefully acknowledged. The authors acknowledge facilities under the National Research Foundation, Prime Minister's Office, Singapore, under its Medium Sized Centre Program, the Singapore University of Technology and Design and the Agency for Science, Technology, and Research.

## Contributions

K. O. Y. K., X. X. C., A. A., and D. T. H. T conceived the idea;  K. O. Y. K. and X. X. C. led the experiments with contributions and suggestions from A. A., G. F. R. C., J. W. C., B.-U. S. and A. C.; the manuscript was written by  X. X. C., K. O. Y. K., and D. T. H. T.; all authors read and contributed to the manuscript; D. T. H. T. supervised the project. X. X. C. and K. O. Y. K. contributed equally.